\documentclass[
    ,final            
  ]
  {aipproc}

\layoutstyle{8x11single}

\begin{document}

\title{New variables for brane-world gravity}


\author{Zolt\'{a}n Kov\'{a}cs }
{address={Departments of Theoretical and Experimental Physics, University of
Szeged, Szeged 6720, D\'{o}m t\'{e}r 9, Hungary},
altaddress={Max-Planck-Institut f\"{u}r Astronomie, K\"{o}nigstuhl 17, D-69117
Heidelberg, Germany}}

\author{L\'{a}szl\'{o} \'{A}. Gergely}
{address={Departments of Theoretical and Experimental Physics, University of
Szeged, Szeged 6720, D\'{o}m t\'{e}r 9, Hungary}}

\begin{abstract}
Geometric variables naturally occurring in a time-like foliation of brane-worlds are introduced. These consist of the induced metric and two sets of lapse functions and shift vectors, supplemented by two sets of tensorial, vectorial and scalar variables arising as projections of the two extrinsic curvatures. A subset of these variables turn out to be dynamical. Brane-world gravitational dynamics is given as the time evolution of these variables. 

\end{abstract}

\maketitle

\section{Introduction}

The novel theory of gravitation known as brane-world is based on the introduction of a fourth non-compact spatial dimension. The fabric of space-time thus becomes a five-dimensional (5d) pseudo-Riemannian manifold (the bulk) in which gravity develops according to the Einstein equation. The particles and fields of the standard model are confined to a four-dimensional (4d) hypersurface, called the brane, kept together by the so-called brane-tension ${}^{(4)}\lambda$. Gravitons are the only ingredients of the model, which can propagate in all dimensions. (Sometimes non-standard model fields, like scalar fields or radiation of quantum origin are also allowed in the bulk.) This new setup gives modifications of the gravitational interaction on the brane at high energies. In a cosmological brane-world scenario, the early universe evolves differently \cite{BDEL}.

The simplest such brane-world, introduced by Randall and Sundrum \cite{RS2}, identifies a flat brane to a domain wall in a 5d anti-de Sitter space-time. They have considered branes with both negative and positive tensons in order to investigate how matter can be localized on the domain wall. For a single brane with positive tension (RS2 model) they have provided a mechanism which confines the matter fields onto the brane. Later on, various generalizations of this model to curved branes have been investigated. In these scenarios the 5d Einstein equations couple the curved geometry of the bulk endowed with a cosmological constant ${}^{(5)}\Lambda$ to the matter fields on the brane (located at $\chi=0$): 
\begin{eqnarray}
{}^{(5)}{G}_{ab}&=&{}^{(5)}\kappa^2\;{}^{(5)}{T}_{ab} \nonumber\\
{}^{(5)}{T}_{ab} &=&-{}^{(5)}\Lambda\;{}^{(5)}g_{ab} + ({}^{(4)}\lambda\; {}^{(4)}g_{ab} + {}^{(4)}T_{ab})\delta(\chi)\ .  \label{G5}
\end{eqnarray}
Here ${}^{(5)}\kappa^2$ is the bulk coupling constant and ${}^{(4)}T_{ab}$ represents the energy-momentum tensor of the standard model fields confined to the brane.

By applying the geometrical approach of the space-time embedding, Shiromizu et al. \cite{SMS} have derived the Einstein equations induced on the brane. They have used the Gauss and Codazzi equations giving the various projections of the 5d geometry onto the brane, which, together with the Lanczos-Sen-Darmoise-Israel matching conditions \cite{Lanczos}-\cite{Israel} and the assumption of $Z_2$ symmetry across the brane, provided the modified version of the 4d field equation, with correction terms as compared to Einstein's gravity:
 \begin{equation}
{}^{(4)}{G}_{ab}= {}^{(4)}\Lambda\; {}^{(4)}g_{ab} + {}^{(4)}\kappa^2\;{}^{(4)}T_{ab} + {}^{(5)}\kappa^2{S}_{ab} -{\mathcal E}_{ab}\ ,  \label{G4}
\end{equation}
The new source terms are of two kinds: contributions from the matter fields and a purely gravitational source from bulk gravity. The new matter source term 
\begin{equation}
S_{ab}=\frac{1}{12}TT_{ab}-\frac{1}{4}T^c{}_aT_{cb}+\frac{1}{24}g_{ab}
(3T_{cd}T^{cd}-T^2)\ 
\end{equation}
is quadratic in the brane energy-momentum tensor.
The second source term specific to brane-world gravity is due to the non-local gravitational field of the bulk. It is the so-called "electric" part 
\begin{equation}
{\mathcal E}_{ab}={}^{(5)}W_{abcd}n^c n^d\ 
\end{equation}
of the 5d bulk Weyl tensor, ${}^{(5)}W_{abcd}$, calculated in the limit $\chi\rightarrow0$.  

The presence of the source term ${\mathcal E}_{ab}$ in the modified Einstein equation (\ref{G4}) indicates that the gravitational equations on the brane do not close. The 4d contracted Bianchi identities imply the constraint
\begin{equation}
{}^{(4)}\nabla^a {\mathcal E}_{ab}={}^{(5)}\kappa^2\;{}^{(4)}\nabla^a S_{ab}\ ,\label{nabEab}
\end{equation}
revealing that ${\mathcal E}_{ab}$ is not freely specifiable but its divergence depends on stress-energy. 
For the full decomposition of the 5d Weyl tensor, its "magnetic part" is also defined,
\begin{equation}
{\mathcal B}_{abc}={}^{(4)}g^d{}_a\;{}^{(4)}g^e{}_b\; n^f\;{}^{(5)}W_{decf}\ ,
\end{equation}
and the effective gravitational equations in the brane-world can be written in terms of the variables ${\mathcal E}_{ab}$ and ${\mathcal B}_{abc}$. These equations are to be solved under the boundary condition (\ref{nabEab}) and a second boundary condition arising from fixing the variable ${\mathcal B}_{abc}$ on the brane \cite{SMS}. 

A covariant decomposition of the new source terms with respect to cosmological symmetries 
can also be given \cite{Maartens}. Then the effective total energy-momentum tensor 
together with conservation theorems are used to provide evolution equations for the local and non-local energy densities and fluxes.

However, an alternative way to deal with the bulk gravitational degrees of freedom is also possible. 
Instead of employing various projections of the 5d Weyl tensor, guided by a geometric motivation \cite{brane_can}, we define here new dynamical variables. These variables are equally adequate to describe bulk gravity and they can be introduced in the decomposition of the effective energy-momentum tensor. Besides, these variables are better suited to develop a canonical theory of brane-world gravity.  

\section{Canonical brane-world gravity}

\subsection{The 3+1+1 decomposition of the bulk}

\begin{figure}
  \includegraphics[height=.4\textheight]{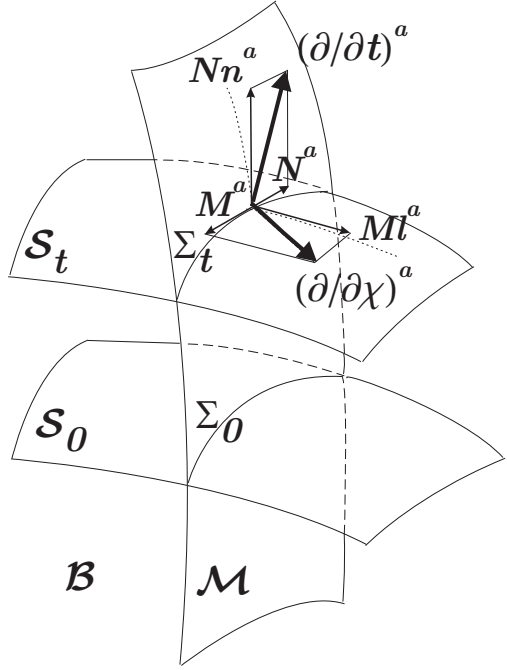}
  \caption{The brane is the hypersurface $\mathcal{M}$ (at $\chi=0$) embedded 
into the bulk $\mathcal{B}$. The foliation of the bulk by spatial hypersurfaces 
is represented by two leaves, taken at $t=0$ and at some generic $t>0$. The leaves 
intersect the brane at $\Sigma_{0}$ and $\Sigma_{t}$, respectively. These 
represent our detectable universe at the two instants of time. The decomposition 
of the time evolution and off-brane evolution vectors is also shown. In order to 
obey the Frobenius theorem, the off-brane component of the time-evolution vector 
was chosen to vanish.}
\label{Figure1}
\end{figure}

The Hamiltonian theory of general relativity is based on the embedding of 3-dimensional (3d) space-like hypersurfaces, representing the space at a given instant, into the 4d space-time manifold. A similar embedding can be constructed in the brane-world scenario as well, by introducing a foliation of the 4d brane (which is already embedded in the 5d bulk). Since the co-dimension of the embedded 3-spaces with respect to the bulk is two, a two-parameter family of the hypersurfaces $\Sigma_{t,\chi}$ ($t,\chi\in I\!\!R$) is required for the 3+1+1 decomposition of the bulk $\mathcal B$. Besides the parameter $t$ representing the many-fingered time in the canonical formalism, here we have introduced the parameter $\chi$. The locus of the brane in the bulk is chosen at $\chi=0$. For any leave $\Sigma_{t}=\Sigma_{t,\chi=0}$ (see Fig. \ref{Figure1} ), representing the conventional 3d space, there are two normal vector fields (one time-like and one space-like), which are independent. We denote them as $n^a$ and $l^a$, with $n^a n_a = -1$, $l^a l_a = 1$ and $n^a l_a = 0$.

Then temporal and off-brane evolutions along vector fields defined by the derivatives of the parameters $t$ and $\chi$ are given as  
\begin{equation}
\left( \frac{\partial }{\partial t}\right) ^{a} = Nn^{a}+N^{a}\ ,  \qquad
\left( \frac{\partial }{\partial \chi }\right) ^{a} = M^{a}+Ml^{a}\ ,
\label{dd}
\end{equation}
where $N^{a}$ and $N$ have the well-known interpretation from the decomposition 
of the $3+1$ dimensional space-time as shift vector and lapse function. 
The vector $M^a$ and the scalar $M$ are the "shift vector" and "lapse function" 
of the off-brane evolution, but as the off-brane evolution vector 
$\partial /\partial \chi $ is space-like rather than time-like, one 
does not expect these quantities to have the same non-dynamical role as $N^{a}$ and $N$. 
Indeed, they characterize the off-brane sector of gravity.
Notably, there is no component along $n^a$ of $\partial /\partial \chi $, 
because the off-brane evolution is chosen along a constant time hypersurface. 
Similarly, there is no component of the shift of $\partial /\partial t $ 
in the off-brane direction, as the two foliations were chosen perpendicularly
\footnote{For generic (non-perpendicular) double foliations the Frobenius theorem, 
stating that the commutator of any pair of tangent vectors of an embedded 
submanifold (here the brane) belongs to the tangent space of that submanifold, 
imposes a constraint on the components of $\partial /\partial \chi $ and 
$\partial /\partial t $ \cite{brane_can}. This constraint is automatically 
satisfied for perpendicular foliations.}. 

The 5d metric defined on $\mathcal{B}$ is decomposed as 
\begin{equation}
{}^{(5)}{g}_{ab}=g_{ab}+l_{a}l_{b}-n_{a}n_{b}\ ,  \label{tildeg0}
\end{equation}
where $g_{ab}$ is the induced metric of $\Sigma _{t}$ with $\sqrt{-{}^{(5)}g}=NM\sqrt{g}$.
The intrinsic geometry of each 3-space, given by $g_{ab}$, describes only the local properties of the brane gravity on the leaves. The non-local bulk effects on the brane are represented by the new gravitational variables $M^a$ and $M$. 
A simple counting shows that the 15 independent components of the 5-metric ${}^{(5)}g_{ab}$ can be replaced by the equivalent set $\{g_{ab},M^{a},M,N^{a},N\} $, for which the restrictions $
g_{ab}n^{a}=g_{ab}l^{a}=M^{a}n_{a}=M^{a}l_{a}=N^{a}n_{a}=N^{a}l_{a}=0$
apply, plus the information that the \textit{evolution of standard-model fields is constrained to a hypersurface}. 

The set of dynamical quantites $\{g_{ab},M^{a},M\}$ can therefore be considered as canonical coordinates of the 5d gravity, while the non-vanishing $N^a$ and $N$ variables play the role of Lagrange multipliers. Then the momenta canonically conjugated to the coordinates can be defined analogously to those in the the standard $3+1$ decomposition, that is via the extrinsic curvature(s) of $\Sigma_{t}$.

\subsection{The fundamental forms and scalars of $\Sigma _{t }$}

The set of the first fundamental form $g_{ab}$, the vector field $M^a$ and the scalar field $M$ are quantities defined on the brane, characterizing brane-world gravitation. 
As there are two independent normal vector fields to the leaves $\Sigma_{t}$, there are also two types of extrinsic curvatures characterizing the embedding. With respect to $\Sigma_{t}$, they further decompose into tensorial, vectorial and scalar quantities. The tensorial quantities derived from the extrinsic curvatures are two kinds of second fundamental forms
\begin{equation}
K_{ab} = g^{c}{}_{a}g^{d}{}_{b}\:{}^{(5)}{\nabla }_{c}n_{d}\ , \qquad
L_{ab} = g^{c}{}_{a}g^{d}{}_{b}\:{}^{(5)}{\nabla }_{c}l_{d}\ .
\label{extr3}
\end{equation}
While the tensor $K_{ab}$ is already well-known from the standard Arnowitt-Deser-Misner (ADM) decomposition of the 4d space-time, the role of the tensor $L_{ab}$ is not well-known. Other projections of the extrinsic curvatures include the normal fundamental forms
\begin{equation}
\mathcal{K}_{a}=g^{b}{}_{a}l^{c}\:{}^{(5)}{\nabla }_{c}n_{b}\ ,   \qquad
\mathcal{L}_{a} =g^{b}{}_{a}n^{c}\:{}^{(5)}{\nabla }_{c}l_{b}   \label{calK1}
\end{equation}
and the normal fundamental scalars
\begin{equation}
\mathcal{K}=l^{a}l^{b}\:{}^{(5)}{\nabla }_{a}n_{b}\ ,   \qquad
\mathcal{L}=n^{a}n^{b}\:{}^{(5)}{\nabla }_{a}l_{b}\ .  \label{calK2}
\end{equation}
The totality of the tensorial, vectorial and scalar quantities introduced above characterizes the embedding of the submanifolds $\Sigma _{t}$ into the bulk. 
Such (or analogous) quantities can be found in the geometric treatments \cite{Schouten} and \cite{Spivak}. 

Let us make a few remarks of the newly introduced quantities.
Since we have chosen perpendicular normal vector fields ($n^a l_a = 0$), the following relation holds for the normal fundamental forms \cite{brane_can}: 
\begin{equation}
\mathcal{K}_{a}=-\mathcal{L}_{a}\ .
\end{equation}
Then the set \{$K_{ab},\mathcal{K}_a,\mathcal{K},L_{ab},\mathcal{L}$\} characterizes the embedding of the 3-spaces $\Sigma _{t}$ into the bulk.
Further, as $n^{a}$ and $l^{a}$ are hypersurface-orthogonal, the tensorial extrinsic curvatures defined above are symmetric. 

By employing the decompositions (\ref{dd}), it is straightforward to derive relations between the tensorial extrinsic curvatures (second fundamental forms) and the time derivatives and off-brane derivatives of the induced metric:
\begin{equation}
\ K_{ab}  =\frac{1}{2N}\left( \frac{\partial }{\partial t}
g_{ab}-2D_{(a}N_{b)}\right) \ ,   \qquad
L_{ab}  = \frac{1}{2M}\left( \frac{\partial }{\partial \chi }
g_{ab}-2D_{(a}M_{b)}\right) \ .  \label{KabLab}
\end{equation}
Here $D_a$ is the derivative compatible with the induced metric $g_{ab}$. In a similar way, the dual of the normal fundamental form can be related to time- and off-brane derivatives of the two kinds of shift vectors: 
\begin{equation}
\mathcal{K}^{a} = \frac{1}{2NM}\left ( \frac{\partial M^{a}}{\partial t}- \frac{\partial N^{a} }{\partial \chi } +M^{b}D_{b}N^{a}- N^{b}D_{b}M^{a} \right )\ .
\label{Kacal}
\end{equation}
Finally, the normal fundamental scalars can be expressed in terms of the time- and $\chi$-derivatives of the two kinds of lapse functions:
\begin{equation}
\mathcal{K} = \frac{1}{MN}\left( \frac{\partial }{\partial t}
M-N^{a}D_{a}M\right) \ , \qquad
\mathcal{L} = -\frac{1}{MN}\left( \frac{\partial }{\partial \chi }
N-M^{a}D_{a}N\right)\ .  \label{KcalLcal}
\end{equation}
The Eqs. (\ref{KabLab}) and (\ref{KcalLcal}) show that the tensor and scalar fields $L_{ab}$ and $\mathcal{L}$ can be expressed as spatial derivatives of the chosen canonical coordinates, whereas the quantities $K_{ab}$, $\mathcal{K}_a$ and $\mathcal{K}$ give the time evolution of $g_{ab}$, $M^a$ and $M$, respectively. Hence the former ones are not independent variables, while the latter are the velocities of the canonical coordinates. 

In a canonical picture the variables $\{g_{ab},M^a,M;K_{ab},\mathcal{K}^a,\mathcal{K}\}$
replace the bulk metric ${}^{(5)}g_{ab}$. In the next section we derive the equations governing the time evolution of these new canonical variables.

\section{The dynamical equations of the canonical variables}

Dynamical evolution follows from variational principles on an action functional containing either the Lagrangian or the Hamiltonian density of the dynamical system. 
Here however we will follow a simpler route to derive these equations, i.e., to
establish the relations among the time derivatives of the
dynamical data $\{g_{ab},M^{a},M\}$ and the extrinsic curvatures $\{K_{ab},
\mathcal{K}_{a},\mathcal{K}\}$.

As the various projections in the decomposition of the 5d Riemann tensor and its contracted quatities describing the bulk geometry contain the Lie derivatives of the velocity variables with respect to the time-like normal vector field $n^a$,
we can find the time derivatives of these variables from pure algebra. In the previous section we have already given the first set of the dynamical equations for $g_{ab}$, $M^a$ and $M$ in Eqs. (\ref{KabLab}), (\ref{Kacal}) and (\ref{KcalLcal}), which can be rewritten as 
\begin{eqnarray}
\frac{\partial g_{ab}}{\partial t}&=& 2NK_{ab}+\pounds_{\mathbf N} g_{ab}\ ,\label{dotgab} \\
\frac{\partial M^{a}}{\partial t} &=& 2MN\mathcal{K}^{a}+\frac{\partial N^{a} }{\partial \chi } + \pounds_{\mathbf N} M^{a}\ , \label{dotMa}\\
\frac{\partial M}{\partial t} &=& MN\mathcal{K}+\pounds_{\mathbf N} M\ .\label{dotM}
\end{eqnarray}

The second set of dynamical equations for $K_{ab}$, $\mathcal{K}^a$  and $\mathcal{K}$ can be obtained from the projections 
$n^a g^b{}_i n^c g^d{}_j\;{}^{(5)}R_{abcd}$, the Codazzi equation containing 
$g^a{}_i l^b\;{}^{(5)}R_{ab}$ and the projection $l^a l^b\;{}^{(5)}R_{ab}$.
In order to obtain the time derivatives, we decompose the vector field $n^a$ in the Lie derivatives cf. Eq. (\ref{dd}). The second set of dynamical equations become: 
\begin{eqnarray}
\frac{\partial K_{ab}}{\partial t} &=&N[ 2K_{ac}K_{b}^{c}-K_{ab} (K+\mathcal{K})+2\mathcal{K}_{a}\mathcal{K}_{b}+g^{c}{}_{a}g^{d}{}_{b}{}^{(5)}\;{R}_{cd}-R_{ab} \nonumber \\
&&+L_{ab}( L-\mathcal{L}) -2L_{ac} L_{b}^{c} +\pounds_{\mathbf{l}} L_{ab}+M^{-1}D_{b}D_{a}M] +D_{b}D_{a}N+\pounds_{\mathbf{N}}K_{ab}\ , \\
\frac{\partial \mathcal{K}_{a}}{\partial t} &=&N[-K\mathcal{K}_{a}+g^b{}_a l^c\;{}^{(5)}R_{bc}-D^{b}L_{ab}+D_{a}
( L-\mathcal{L})]-( L_{a}^{b}+\mathcal{L}\delta _{a}^{b}) D_{b}N   +\pounds_{\mathbf{N}}\mathcal{K}_{a}\ , \\
\frac{\partial }{\partial t}\mathcal{K} &=&N[
-2\mathcal{K}_{a}\mathcal{K}^{a}-\mathcal{K}( K+\mathcal{K})
l^{a}l^{b}{}^{(5)}{R}_{ab}-L_{ab}L^{ab}+\mathcal{L}^{2} \nonumber\\
&&+M^{-1}D_{a}D^{a}M+ \mathcal{L}_{\mathbf{l}}\left( L-\mathcal{L}\right) ]
+M^{-1}D^{a}MD_{a}N+\pounds_{\mathbf{N}}\mathcal{K}\ \label{dotKcal}.
\end{eqnarray}
The projections of the 5d Riemann and Ricci tensors appearing in the above equations are determined from the bulk Einstein equation (\ref{G5}). Whenever the bulk contains nothing but a cosmological constant, however the matter on the brane is left arbitrary, the projection $g^b{}_a l^c\;{}^{(5)}R_{bc}$ vanishes, as proven in \cite{brane_can}.
Such a matter source can be decomposed as follows:
\begin{eqnarray}
{}^{(5)}{T}_{ab} &=&-{}^{(5)}{\kappa }^{2}\;{}^{(5)}{\Lambda }l_{a}l_{b}+
\left[ \left( \rho +\lambda \right) \delta \left( \chi \right) +{}^{(5)}{
\kappa }^{2}{}^{(5)}{\Lambda }\right] n_{a}n_{b}\nonumber\\
&&+\left[ \left( p-\lambda \right) \delta \left( \chi \right) -{}^{(5)}{
\kappa }^{2}\;{}^{(5)}{\Lambda }\right] g_{ab} +\left[ \Pi _{ab}+2n_{(a}Q_{b)}\right] \delta \left( \chi \right) \ ,
\label{tildeT}
\end{eqnarray}
with energy density $\rho $, homogeneous pressure $p$, tensor of anisotropic pressures $\Pi _{ab}$ and energy transport $Q_{a}$, all measured by observers moving along the congruence $n^a$. 

Thus the equations (\ref{dotgab})-(\ref{dotKcal}) completely determine the dynamical evolution of the brane-world gravity in terms of the newly introduced, geometric variables.

\section{Summary}

We have presented here the main results derived in detail in \cite{brane_can}, which rely on the  $3+1+1$ decomposition of a 5d space-time, both with respect to a timelike and a spacelike direction. 

The bulk metric is replaced by two sets of variables. The first
set consist of one tensorial (the induced metric $g_{ab}$), one vectorial ($M^{a}$) and one scalar ($M$) dynamical quantity, all defined on the 3-space. Their time evolutions are related to the second fundamental form
(the extrinsic curvature $K_{ab}$), the normal fundamental form ($\mathcal{K}
^{a}$) and normal fundamental scalar ($\mathcal{K}$), respectively. The
non-dynamical set of variables is given by the lapse function and the shift
vector, which however has one component less. The missing off-brane component is due
to the externally imposed constraint, stating that physical
trajectories are confined to the 4d brane. The pair of dynamical variables ($g_{ab}$, $K_{ab}$), well-known from the ADM
decomposition is supplemented by the pairs ($M^{a}$, $\mathcal{K}^{a}$) and (
$M$, $\mathcal{K}$) representing the bulk degrees of freedom. 

We have given the set of equations characterizing gravitational dynamics on a brane with arbitrary matter embedded into a bulk with cosmological constant.
As a completion of our program of characterizing brane-world gravity in terms of geometric variables, the electric part of the Weyl tensor should be given in terms of these variables. This is work in progress \cite{brane_can2}.

Our formalism is suited for the study of the initial
value problem and for canonical gravitational dynamics in generalized RS2 brane-world
scenarios. 

\begin{theacknowledgments}
This work was supported by OTKA grants no. T046939 and TS044665. L.\'A.G.
was further supported by the J\'{a}nos Bolyai Scholarship of the Hungarian
Academy of Sciences.
\end{theacknowledgments}

\bibliographystyle{aipproc}   
\bibliography{new_variables_kovacs_gergely.bib}

\IfFileExists{\jobname.bbl}{}
 {\typeout{}
  \typeout{******************************************}
  \typeout{** Please run "bibtex \jobname" to optain}
  \typeout{** the bibliography and then re-run LaTeX}
  \typeout{** twice to fix the references!}
  \typeout{******************************************}
  \typeout{}
 }

\end{document}